\documentclass[a4paper,11pt]{article}
\usepackage{jcappub} 
\usepackage{lineno}

\arxivnumber{1234.56789} 
\title{\boldmath Playground of Lognormal Seminumerical Simulations of~the~Lyman~$\alpha$ Forest: Thermal History of the Intergalactic Medium}

\author[a,b,1]{T. Ondro%
\note{Corresponding author.}}
\author[c]{B. Arya}
\author[d]{R. G\'{a}lis}
\affiliation[a]{Department of Technology and Automobile Transport, Faculty of AgriSciences, Mendel University in Brno, Zem\v{e}d\v{e}lsk\'{a} 1, \\ 613 00 Brno, Czech Republic}
\affiliation[b]{Department of Applied Mathematics, VSB - Technical University of Ostrava, 17. listopadu 15/2172, 70800 Ostrava, Czech Republic}
\affiliation[c]{Department of Space, Planetary \& Astronomical Sciences \& Engineering, Indian Institute of Technology, Kanpur 208016, Uttar Pradesh, India}
\affiliation[d]{Institute of Physics, Faculty of Science, Pavol Jozef \v{S}af\'{a}rik University in Ko\v{s}ice, Park Angelinum 9,
040 01 Ko\v{s}ice, Slovakia}

\emailAdd{tomas.ondro@mendelu.cz}
\emailAdd{bhaskara@iitk.ac.in}
\emailAdd{rudolf.galis@upjs.sk}

\abstract{This study aims to test a potential application of lognormal seminumerical simulations to recover the thermal parameters and Jeans length. This could be suitable for generating large number of synthetic spectra with various input data and parameters, and thus ideal for interpreting the high-quality data obtained from QSO absorption spectra surveys. We use a seminumerical approach to simulate absorption spectra of quasars at redshifts $ 3 \leq z \leq 5$. These synthetic spectra are compared with the 1D flux power spectra and using the Markov Chain Monte Carlo analysis method we determine the temperature at mean density, slope of the temperature-density relation and Jeans length. Our best-fit model is also compared with the evolution of the temperature of the intergalactic medium from various UVB models. We show that the lognormal simulations can effectively recover thermal parameters and Jeans length. Besides, by comparing the synthetic flux power spectra with observations from Baryon Oscillation Spectroscopy Survey we found, that such an approach can be also used for the cosmological parameter inference.}

\begin{document}
\maketitle
\flushbottom

\section{Introduction}
The primary observational evidence for the intergalactic medium (IGM) comes from absorption lines at wavelengths shortward of the Lyman-$\alpha$ (Ly-$\alpha$) emission line. These are thought to be the Ly-$\alpha$ absorption lines arising from the moderate overdensities ($\Delta \lesssim 10$) in the gas density field \citep{viel+02, Viel2002, demi_2011}. Thus, these absorption lines in the spectra of high redshift sources are being revealed as an extremely powerful tool in observational cosmology \citep{FontRibera2012}. For example, the final sample from the extended Baryon Oscillation Spectroscopy Survey (eBOSS) from \cite{Chabanier2019} contains approximately 210,000 quasars \citep{Bourboux2020}. This imposes a condition to produce computationally inexpensive, but realistic datasets. For this purpose, it is therefore suitable to use the numerical and semianalytical models for the IGM, which are developing from the 1990s \citep{Bi1992,Bi1993,Bi1995,Bi1997}. Although this is relatively simple model, it can predict various properties of the absorption lines, such as column density distribution and the distribution of linewidths \citep{Viel2002}.

Nowadays, a majority of studies which are focused on the statistical properties of the Ly-$\alpha$ forest use modern hydrodynamical simulations for interpreting the observations. These simulations are, on the one hand, more accurate and include most of the relevant physics to model the Ly-$\alpha$ forest. On the other hand, they are computationally extensive which is the limitation in the case, when we want to explore large parameter space. Therefore, the efficient semi-numerical models of the Ly-$\alpha$ forest may have an important role in interpreting of the cosmological and astrophysical parameters from QSO absorption spectra surveys.

In the past, many such semi-numerical methods have been proposed, such as: (i) assuming baryons trace a smoothened dark matter density field \citep{petitjean_1995, croft2002}, (ii) Taylor expanding the observables around a "best-guess" model interpreted from few hydrodynamic simulations, (iii) semi-analytical modeling using lognormal approximation \citep{cj91, HuiGnedin, Bi1997, Choudhury2001, Arya2023, Arya2024}. Results from these methods show that the Ly-$\alpha$ forest is sensitive to the astrophysical processes in the IGM, thereby marginalizing these astrophysical parameters is imperative to constraining cosmology and dark matter models using the Ly-$\alpha$ forest.

The aim of this study is to test the potential application of lognormal seminumerical simulations to recover the temperature at mean density, slope of the temperature-density relation and Jeans length between $3\leq z \leq 5$. The lognormal allows us to efficiently generate Ly-$\alpha$ spectra while accounting for non-Gaussianities in the baryonic density field to a certain extent. We utilise the Ly-$\alpha$ flux power spectrum, which is often used to constrain the thermal \citep{Boera2019, Walther2019,Gaikwad2021} and also the cosmological parameters \citep{McDonald2000,Viel2004}, to compare the observational results with our model. 

The article is organised as follows: Section \ref{Sec:Theor_bkg} contains the theoretical background of the lognormal seminumerical simulations of the Ly-$\alpha$ forest. Description of the Ly-$\alpha$ flux power spectrum calculation is presented in Section \ref{Sec:FPS}. These two sections is meant to provide a comprehensive description of the methods used in this study, making them fairly technical in nature. Section \ref{Sec:Obs_data} contains information about observational data and methodology for the Markov Chain Monte Carlo inference used to restore the thermal parameters and Jeans length. In Section \ref{Sec:Results}, we presents our results for the best-fit model and the comparison of determined values of temperature at a mean density with the thermal evolution of the IGM. Our conclusions are given in Section \ref{Sec:Conclusions}.

In this study, we used the default cosmological parameters: $(\Omega_{\Lambda}, \Omega_{m}, \Omega_{b}, \sigma_{8}, n_{s}, h, Y)= (0.692, 0.308, 0.0482, 0.829, 0.961,$ $ 0.678, 0.24)$ consistent with a flat $\Lambda$CDM cosmology \citep{Planck2014}.

\section{ Simulating the QSO absorption spectra }
\label{Sec:Theor_bkg}
In this section, we describe the structure of the lognormal seminumerical model, which is based on previous studies \citep{Choudhury2001,Arya2023,Arya2024}. Note that we also provide algorithms for most tricky parts.

\begin{figure}[ht]
    \centering
    \includegraphics[width=0.75\linewidth]{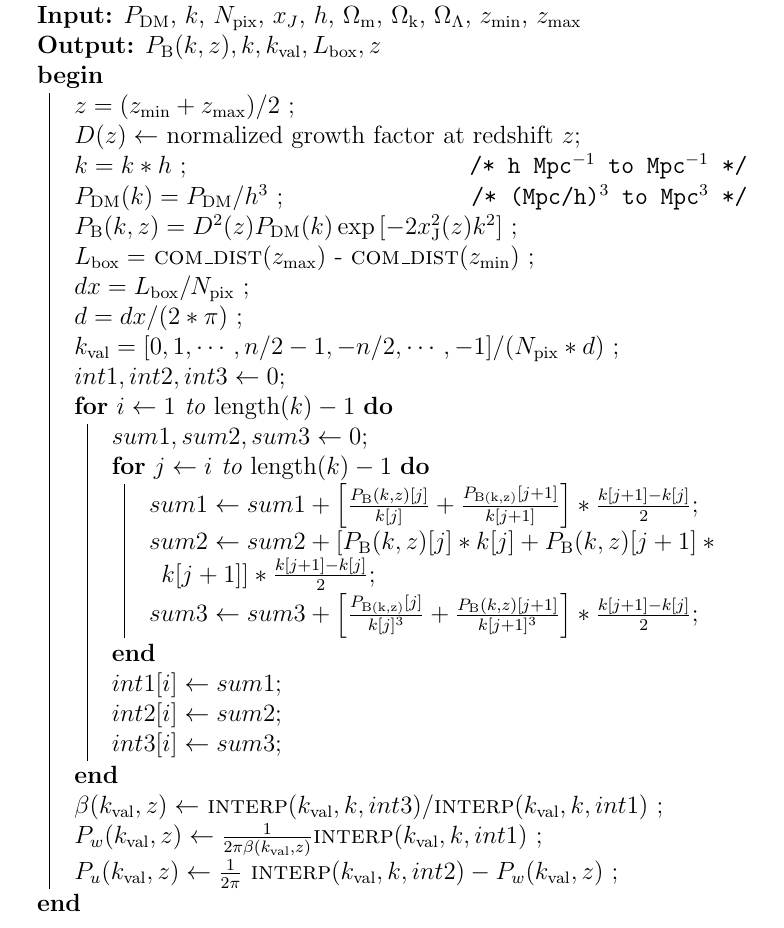}
    \caption{Algorithm for calculating $P_{u}(k,z)$ and $P_{w}(k,z)$. Note that the {\sc com\_dist} and {\sc interp} functions are for the calculation of comoving distance and one-dimensional linear interpolation, respectively.}
    \label{Algorithm:step1}
\end{figure}

We started with linearly extrapolated power spectrum of dark matter (DM) density field $P_{\rm{DM}}(k)$ calculated for a given set of cosmological parameters at the present epoch ($z = 0$). In this study, we used the power spectrum as computed by the {\sc camb} Boltzmann solver \citep{Lewis2000}. Then, the 3D power spectrum of the baryonic density fluctuations for any arbitrary $z$ is given by 
\begin{equation}
    P_{\rm{B}}(k,z) = D^{2}(z) P_{\rm{DM}}(k) \exp{\left[ -2 x_{\rm{J}}^{2}(z) k^{2}\right] },
\end{equation}
where $D(z)$ is the linear growth factor and $x_{\rm{J}}(z)$ is Jeans length. The algorithm for calculating the aforementioned quantities is listed in  Fig. \ref{Algorithm:step1}.

\begin{figure}[h]
    \centering
    \includegraphics[width=0.75\linewidth]{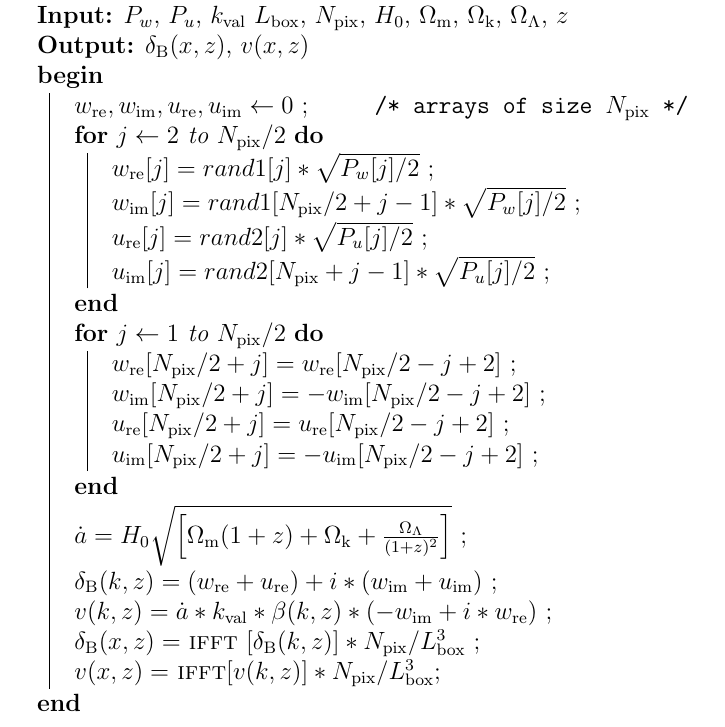}
    \caption{Algorithm for calculating $\delta(x,z)$ and $v(x,z)$. Note that $rand1$ and $rand2$ are random samples of size $N_{\rm pix}$ from Gaussian distribution of zero mean and unit variance, $i$ is the imaginary unit ($i = \sqrt{-1}$) and the function {\sc ifft} computes the one-dimensional inverse discrete Fourier transform.}
    \label{Algorithm:FFT_part}
\end{figure}

To simulate the one dimensional density and velocity fields for a chosen redshift $z$, we start with two independent Gaussian fields $w_{0}(k)$ and $u_{0}(k)$ with unit power spectrum. Then, we prepare two independent Gaussian fields with power spectra $P_{w}(k,z)$ and $P_{u}(k,z)$ as
\begin{equation}
    w(k,z) = w_{0}(k) \sqrt{ P_{w}(k,z) }
\end{equation}
and
\begin{equation}
    u(k,z) = u_{0}(k) \sqrt{ P_{u}(k,z) },
\end{equation}
where
\begin{equation}
    P_{w}(k,z) = \beta^{-1}(k,z) \frac{1}{2 \pi} \int^{\infty}_{\mid k \mid} \frac{{\rm d}k'}{k'} P_{B}(k',z),
    \label{Eq:Pw}
\end{equation}
\begin{equation}
    P_{u}(k,z) = \frac{1}{2 \pi} \int^{\infty}_{\mid k \mid} {{\rm d}}k' k' P_{B} (k',z) - P_{w}(k,z),
    \label{Eq:Pu}
\end{equation}
and
\begin{equation}
    \beta(k,z) = \frac{ \int^{\infty}_{\mid k \mid} ({\rm d}k'/k'^{3}) P_{B}(k',z) }{ \int^{\infty}_{\mid k \mid} ({\rm d}k / k') P_{B}(k',z) }.
\end{equation}
Then, the linear density and velocity fields in the $k$-space are given by
\begin{equation}
    \delta_{\rm B}(k,z) = w(k,z) + u(k,z),
\end{equation}
\begin{equation}
    v(k,z) = i \dot{a} k \beta(k,z) w(k,z),
\end{equation}
where $\dot{a}$ is the time derivative of the scale factor
\begin{equation}
    \dot{a}^{2}(z) = H_{0}^{2} \left[ \Omega_{\rm m}(1 + z) + \Omega_{\rm k} + \frac{\Omega_{\Lambda}}{(1 + z)^{2}} \right] ,
\end{equation}
where $H_{0}$ is the present expansion rate of the Universe, and $\Omega_{\rm m}$, $\Omega_{\rm k}$ and $\Omega_{\Lambda}$ are the cosmological parameters  corresponding to the content of matter, curvature and dark energy, respectively. The $\delta_{\rm B}(x,z)$ and $v(x,z)$ in real comoving space are obtained  using Fourier transforms. The algorithm for calculating aforementioned quantities is listed in Fig. \ref{Algorithm:FFT_part}.

Note that we firstly calculated the Eqs. (\ref{Eq:Pw}) and (\ref{Eq:Pu}) from the all values of 3D power spectrum of the baryonic density fluctuations $P_{\rm B}$, and then we used interpolation at specific query points as we need the values of $P_{w}$ and $P_{u}$ at $k_{\rm vals}$ (see Fig. \ref{Algorithm:step1}).

To account the effect of non-linearities, we assuming the number density distribution of baryons $n_{\rm B}$ to be a lognormal field
\begin{equation}
    n_{\rm B} = A \exp{[\delta_{\rm B}(x,z)]},
    \label{Eq:numb_dens_baryons}
\end{equation}
where $\delta_{\rm B}(x,z)$ is the linear density contrast in baryons. Normalization constant $A$ can be calculated as
\begin{equation}
    A = \frac{n_{0}(z)}{\langle \exp{[\delta_{\rm B}(x,z)]} \rangle},
    \label{Eq:Norm_constant}
\end{equation}
where 
\begin{equation}
    n_{0} = \frac{\Omega_{\rm B} \rho_{\rm C}}{\mu_{\rm B} m_{\rm p}} (1 + z)^{3},
\end{equation}
where $\rho_{\rm C}$ corresponds to the critical density of Universe, $\Omega_{\rm B}$ is the baryonic density parameter and the term in denominator corresponds to the mass per baryonic particle given by $\mu_{\rm B} m_{\rm p} = 4 m_{\rm p} / (4 - 3Y)$. Combining equations (\ref{Eq:numb_dens_baryons}) and (\ref{Eq:Norm_constant}) leads to
\begin{equation}
    n_{\rm B}(x,z) = n_{0}(z) \frac{\exp{[\delta_{\rm B}(x,z)}]}{\langle \exp{[\delta_{\rm B}(x,z)]} \rangle}.
\end{equation}
Under the assumption of photoionization equilibrium,
\begin{equation}
    n_{\text{H\,\textsc{\lowercase{i}}}} = \frac{\alpha(T) n_{\rm p} n_{\rm e}}{\Gamma_{\text{H\,\textsc{\lowercase{i}}}}}(z),
\end{equation}
where $\alpha(T)$ is the recombination coefficient at temperature $T$, $n_{\rm p}$, $n_{\rm e}$ are the number densities of protons and free electrons, respectively, and $\Gamma_{\text{H\,\textsc{\lowercase{i}}}}$ is the hydrogen photoionization rate. Assuming a fully ionized IGM, number densities of protons and free electrons are given by
\begin{equation}
    n_{\rm p}(x,z) = \frac{4(1-Y)}{4 - 3Y} n_{\rm B}(x,z)      
\end{equation}
and
\begin{equation}
    n_{\rm e}(x,z) = \frac{4 - 2Y}{4 - 3Y} n_{\rm B}(x,z),
\end{equation}
where $Y$ is the helium weight fraction. Note that we take the temperature dependence of $\alpha(T)$ to be given by \cite{Rauch1997}
\begin{equation}
    \alpha(T) = 4.2 \times 10^{-13} \left[ \frac{T(x,z)}{10^{4}\rm{K}} \right] ^{-0.7}.
\end{equation}
The temperature field can be related to the baryonic density using the following equation
\begin{equation}
    T(x,z) = T_{0}(z) \left( \frac{n_{\rm b}(x,z)}{n_{0}(z)} \right) ^{\gamma(z) - 1},
\end{equation}
where $T_{0}$ is the temperature at the mean density and $(\gamma - 1)$ is a~power-law index. 

\begin{figure}
    \centering
    \includegraphics[width=0.5\linewidth]{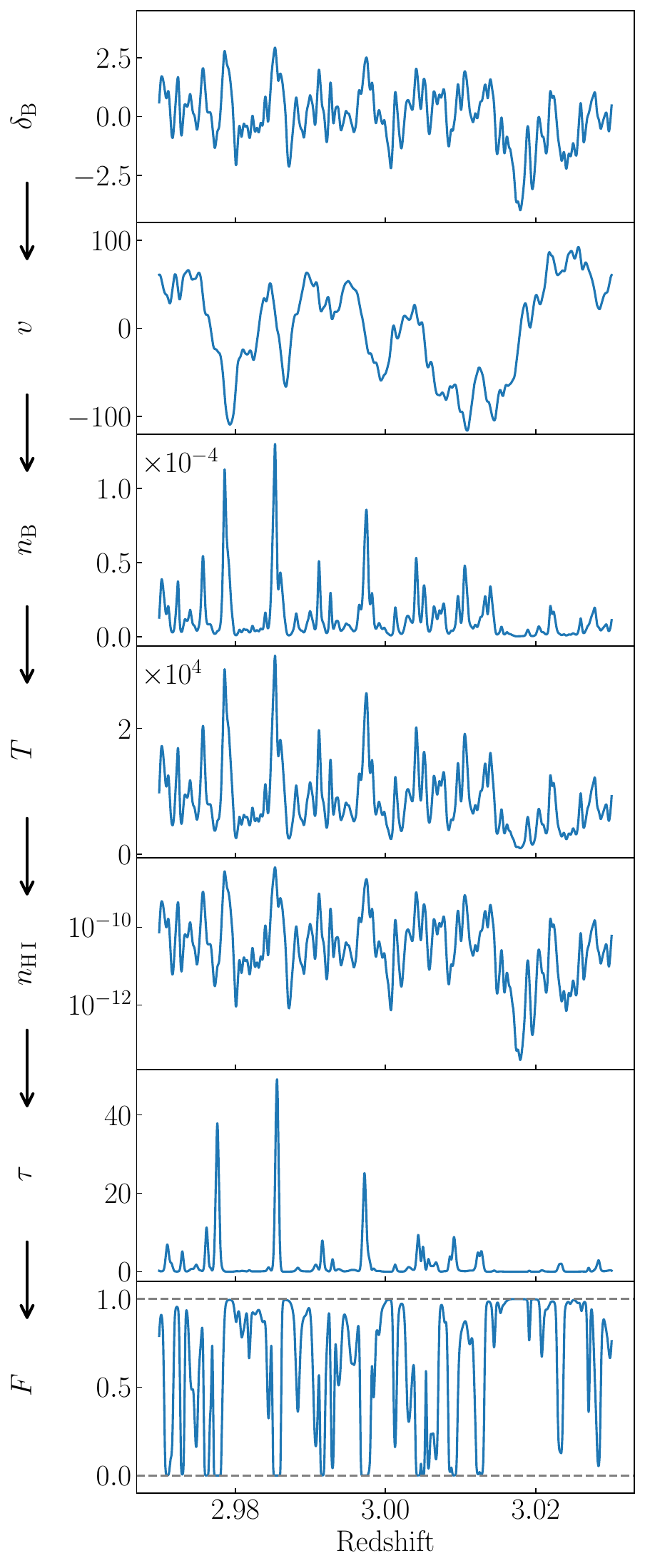}
    \caption{Sample of one line of sight at different stages of transformation.}
    \label{fig:mock_skewer}
\end{figure}

The last step is to calculate the transmitted flux according to the equation
\begin{equation}
    F = F_{\rm c} \exp{(-\tau)},
\end{equation}
where continuum flux $F_{\rm c}$ was set up to unity. The Ly-$\alpha$ optical depth is calculated as
\begin{equation}
\begin{split}
    \tau(x_i,z) = &\frac{c I_{\alpha}}{\sqrt{\pi}} \sum_{j} \delta x \frac{n_{\text{H\,\textsc{\lowercase{i}}}}(x_{j},z)}{b(x_{j},z)[1 + z(x_{j})]}\\ &\times V_{\alpha} \left( \frac{c[z(x_{j} - z(x_{i}))]}{b(x_{j},z)[1 + z(x_{i})]} + \frac{v_{b}[x_{j},z]}{b(x_{j},z)} \right),
\end{split}
\end{equation}
where $i,j$ corresponds to the indexes of the points along the line of sight (LOS), $I_{\alpha}$ is the Ly-$\alpha$ absorption cross-section, $\delta x$ is the separation between the grid points, and $V_{\alpha}$ is the Voigt profile for the Ly-$\alpha$ transition. The quantity 
\begin{equation}
    b(x,z) = \sqrt{ \frac{2 k_{\rm b} T(x,z)}{m_{\rm p}} }
\end{equation}
is the Doppler parameter and $k_{\rm b}$ is Boltzmann constant. We also plot steps of transformation in Fig. \ref{fig:mock_skewer}. 
%
\section{Ly-\texorpdfstring{$\alpha$}{alpha} flux power spectrum}
\label{Sec:FPS}
The Ly-$\alpha$ flux power spectrum is often used to constrain the thermal \citep{Boera2019, Walther2019,Gaikwad2021} and also the cosmological parameters \citep{McDonald2000,Viel2004}. In this work, we performed the power spectrum measurements on the flux contrast estimator
\begin{equation}
    \delta_{\rm F} = \frac{F(v) - \left\langle F \right\rangle}{\left\langle F \right\rangle},
    \label{Eq:contrast_estimator}
\end{equation}
where $F(v)$ is the transmission in the Ly-$\alpha$ forest and $\left\langle F \right\rangle$ is the mean flux. The transmitted flux power spectrum (FPS) is computed according \cite{Villasenor2021} as
\begin{equation}
    P(k) = v_{\rm max} \langle \mid \tilde{\delta}_{\rm F}(k) \mid ^2 \rangle, 
\end{equation}
where
\begin{equation}
    \tilde{\delta}_{\rm F}(k) = \frac{1}{v_{\rm max}} \int_{0}^{v_{\rm max}} \exp{(-iku)} \delta_{\rm F}(v)\,{\text d}v.
    \label{Eq:fourier_modes}
\end{equation}
The FPS is often expressed in terms of the dimensionless quantity
\begin{equation}
    \Delta^{2}_{\rm F}(k) = \frac{1}{\pi} k P(k),
\end{equation}
where $k = 2 \pi / v$.
\begin{figure}
    \centering
    \includegraphics[width=0.5\linewidth]{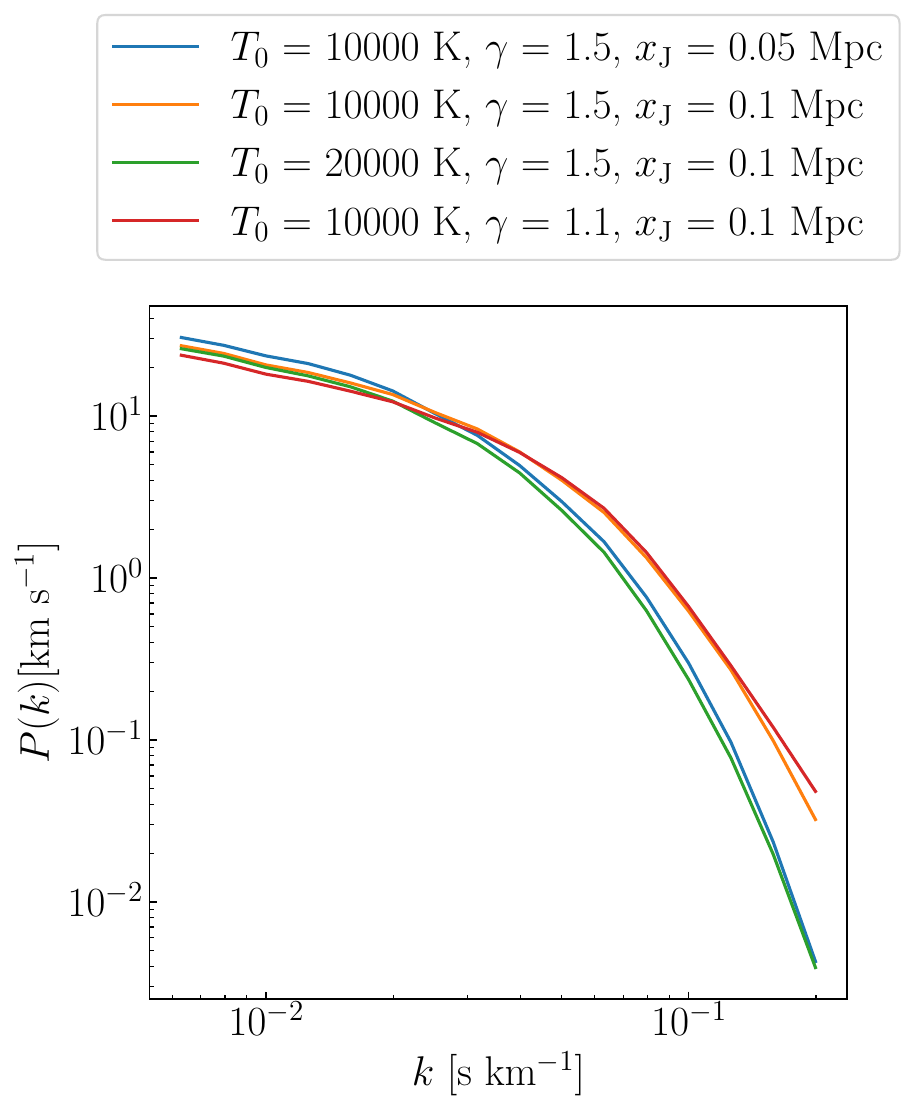}
    \caption{Sensitivity of FPS to the parameters $T_{0}$, $\gamma$ and $x_{J}$ for four models.}
    \label{fig:comp_thermal}
\end{figure}
The sensitivity of FPS to the parameters $T_{0}$, $\gamma$ and $x_{J}$ is shown in Fig.\ref{fig:comp_thermal}. The FPS in our models at scales $k < 0.01$ is nearly the same because of the dark matter density field is the same for all models. It can be also observed that the flux power spectrum is systematically lower for model with higher $T_{0}$ at scales $0.03 < k < 0.2$, whereas is lower for models with higher value of $\gamma$.

\section{Observational Data}
\label{Sec:Obs_data}
In this study, we used the observational determinations of the FPS measured from the studies \cite{Boera2019} and \cite{Irsic2017} for comparison with our models. Note that we also plot the FPS based on the eBOSS measurements \citep{Chabanier2019}. Since this FPS is estimated in the redshift range of $2.2 < z < 4.6$ and probes mostly large scales (corresponds to the interval 0.001 s km$^{-1}$ $\leq k \leq$ 0.02 s km$^{-1}$), we can also test the use of this approach for cosmological parameter inference.

For completeness we add that we do not use the results from the study \cite{Walther2018} because, as was noted in \cite{Villasenor2022}, the estimates show significant differences with those from eBOSS in the overlapping range of scales (0.003~s~km$^{-1} \leq k \leq$ 0.02 s km$^{-1}$).
To get closer to the cosmological simulations, in the case of each redshift bin, we used the 1D array of length 40 $h^{-1}$Mpc with 2\,048 elements, which is suitable for studying the small scale structures probed by the Ly-$\alpha$ forest. 

We applied the MCMC sampler \citep{Lewis2002,Lewis2013,Torrado2019,Torrado2021} to compare the simulated $P(k)$ to the observational measurements in the redshift range of $3 \leq z \leq 5$. To determine when a chain is converged, we use Gelman-Rubin statistics parameter, $R-1 < 0.05$ \citep{Gelman1992}.

In the case of thermal parameters, we adopt flat priors on temperature at the mean density $T_{0}$, $\gamma$ and $x_{J}$ in the ranges $3\,000\,{\rm K} < T_{0} < 20\,000\,{\rm K}$, $0.75 < \gamma < 2.25$, and $0.04\,{\rm Mpc} < x_{J} < 0.4\,{\rm Mpc}$, respectively. For the hydrogen photoionization rate $\Gamma_{\text{H\,\textsc{\lowercase{i}}}}$, we used values from \cite{FG2020}. The likelihood function for the model given by the parameters $\Theta = \{T_{0}, \gamma, x_{\rm J}\}$ is evaluated as
\begin{equation}
    \ln{\mathcal{L}} (\Theta) = - \frac{1}{2} \sum_{k} \left[ \left( \frac{P(k) - P_{\rm model}(k,\Theta)}{\sigma(k)} \right)^{2} + \ln{2 \pi \sigma(k)^{2}}  \right].   
\end{equation}
%
%
\begin{figure}
    \centering
    \includegraphics[width=0.5\linewidth]{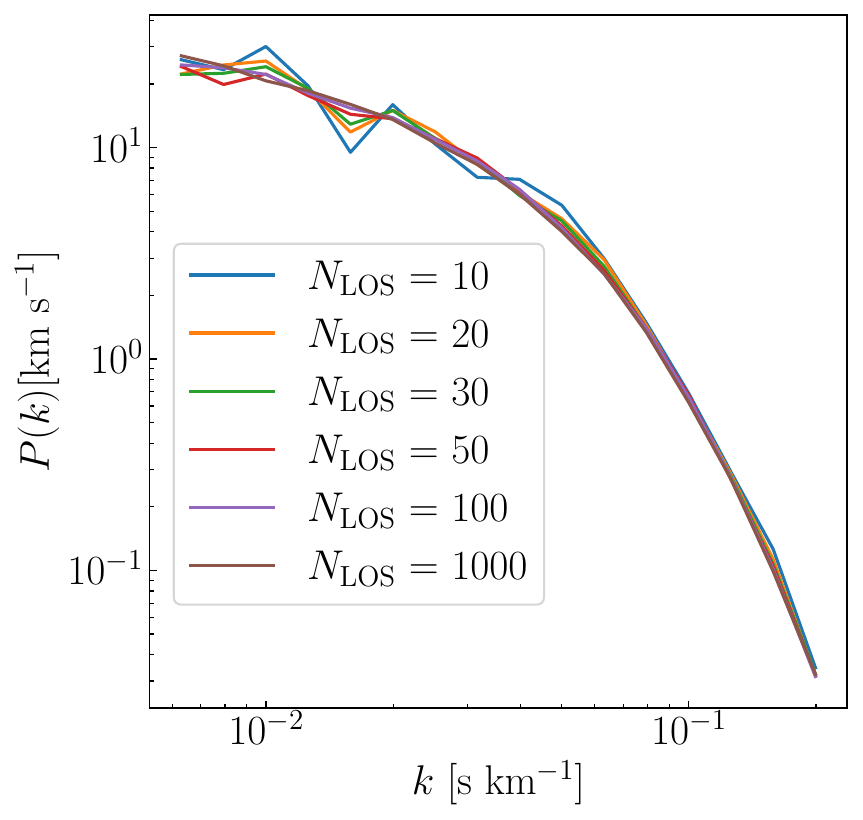}
    \caption{Illustration of the impact of various LOS number used for the FPS calculation. In this case we used the parameters $T_{0} = 10\,000$ K, $\gamma = 1.5$ and $x_{\rm J} = 0.1$ Mpc.}
    \label{fig:convergence_test}
\end{figure}

\begin{figure*}
    \centering
    \includegraphics[width=0.9\linewidth]{figs/Pk_z.pdf}
    \caption{$P(k)$ from the study \cite{Chabanier2019} (red points) and \cite{Boera2019} (blue points), which were used to determine the parameters $T_{0}, \gamma$ and $x_{J}$. The best-fit evolutions of $P(k)$ based on the $N_{\rm LOS} = 30$ and $N_{\rm LOS} = 1000$ are shown with green and orange curves, respectively. The fractional differences of the observations and the best-fit model based on $N_{\rm LOS} = 30$ are shown in the bottom part of each panel. Note that the differences from the eBoss measurements are not plotted.}
    \label{fig:Pk_evolution}
\end{figure*}
\section{Results and Discussion}
\label{Sec:Results}
\subsection{\texorpdfstring{$P(k)$}{P(k)} Model Comparison with Data}
The comparison of the best-fit synthetic FPS at various redshifts is shown in Fig.\ref{fig:Pk_evolution} and the parameter estimates obtained from the MCMC runs are presented in Tab. \ref{Tab:results} and Appendix \ref{Appendix:MCMC}.

\begin{table}[h]
\caption{Values of the determined parameters}              
\label{Tab:results}
\centering
\begin{tabular}{c c c c}         
\hline\hline                        
redshift & $T_{0}$ [K] & $\gamma$ & $x_{J}$ [Mpc] \\    
\hline                                   
    \vspace{0.1 cm}
    3.0 & $16\,746^{+1645}_{-1763}$ & $1.42^{+0.20}_{-0.26}$ & $0.129^{+0.017}_{-0.019}$ \\
    \vspace{0.1 cm}
    3.2 & $17\,858^{+1395}_{-1893}$ & $1.40^{+0.09}_{-0.09}$ & $0.085^{+0.015}_{-0.011}$ \\
    \vspace{0.1 cm}
    3.4 & $15\,609^{+1301}_{-1236}$ & $1.53^{+0.07}_{-0.07}$ & $0.079^{+0.012}_{-0.011}$ \\
    \vspace{0.1 cm}
    3.6 & $18\,115^{+1151}_{-1322}$ & $1.42^{+0.08}_{-0.08}$ & $0.072^{+0.010}_{-0.009}$ \\
    \vspace{0.1 cm}
    3.8 & $18\,163^{+1267}_{-3055}$ & $1.44^{+0.15}_{-0.37}$ & $0.100^{+0.012}_{-0.008}$ \\
    \vspace{0.1 cm}
    4.0 & $16\,375^{+2463}_{-2443}$ & $1.15^{+0.24}_{-0.26}$ & $0.094^{+0.008}_{-0.009}$ \\
    \vspace{0.1 cm}
    4.2 & $8\,370^{+2064}_{-1589}$ & $1.16^{+0.23}_{-0.21}$ & $0.071^{+0.006}_{-0.008}$ \\
    \vspace{0.1 cm}
    4.6 & $8\,855^{+1756}_{-1461}$ & $1.20^{+0.17}_{-0.15}$ & $0.061^{+0.003}_{-0.004}$ \\
    \vspace{0.1 cm}
    5.0 & $10\,870^{+2017}_{-2267}$ & $1.19^{+0.18}_{-0.20}$ & $0.047^{+0.004}_{-0.004}$ \\
\hline                                             
\end{tabular}
\end{table}

In general, the synthetic FPS are consistent on large and intermediate scales (0.003 s km$^{-1}$ $\leq k \leq$ 0.04 s km$^{-1}$) with the estimates from \cite{Irsic2017}, with minor differences on large scales in the case of $z = 3$ and $z = 3.2$. However, there are systematic discrepancies on the scales $k > 0.04$, where the models has less power than that from \cite{Irsic2017}. This behaviour may suggest, that for the correct reproduction of the FPS it is necessary to use dataset, which coverings smaller scales.
 
On the other hand, the results show, that synthetic flux power spectrum is in good agreement with the high redshift measurements of $P(k)$ from the study \cite{Boera2019}. Also, the best-fit synthetic FPS shows good agreement with the large scale $P(k)$ with the results from eBOSS except of the redshifts $z = 3.2$, $z = 3.8$ and $z = 4.0$. Thus, such an approach can be also used for the cosmological parameter inference.

There is also limitation of lognormal seminumerical model, which is important to realize. The Jeans length depends on temperature and density, and therefore should be adaptive \citep{Viel2002}. For the simplicity in this case, and also in most practical implementations, it is assumed to be a constant. However, this leads to the equal smoothing of all dark matter structures independent of their density.

By comparing the lognormal model with the smooth particle hydrodynamical simulations was found that the lognormal model cannot simultaneously recover the true value of all parameters. Since this failure mostly manifests in the poor recovery of $\Gamma_{\text{H\,\textsc{\lowercase{i}}}}$, we use only 3-parameter fit \cite{Arya2024}.

\begin{figure}
    \centering
    \includegraphics[width=\linewidth]{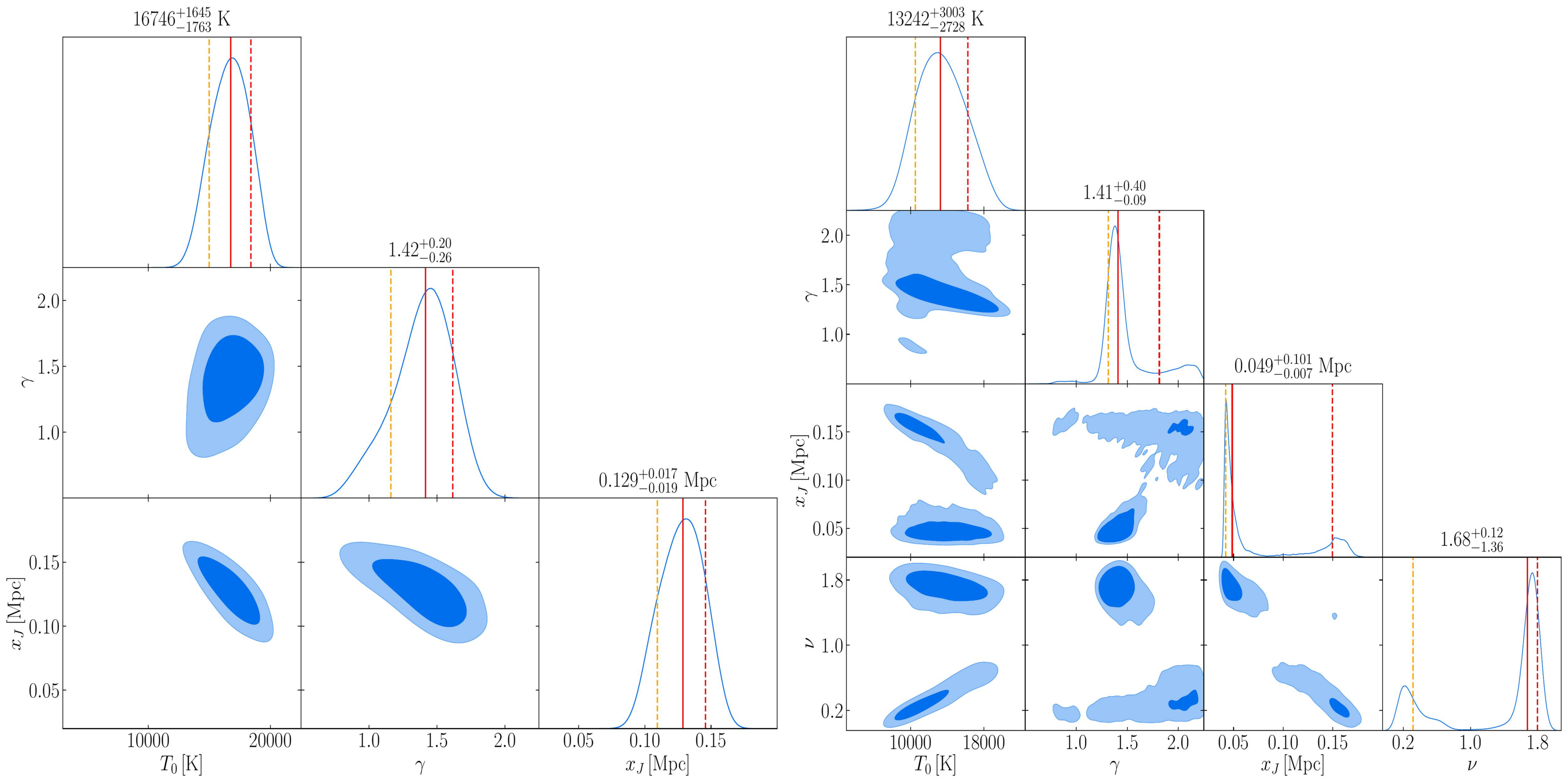}
    \caption{Parameter estimates obtained from the MCMC runs in the 3D (top panel) and 4D (bottom panel) space for redshift $z = 3$. Red solid line corresponds to the median value and orange and red dashed lines correspond to the 16th and 84th percentile, respectively.}
    \label{fig:params_comp}
\end{figure}

Note that in \cite{Arya2024} was stated that four-parameter model $\{ T_{0}, \gamma, x_{\rm J}, \Gamma_{\text{H\,\textsc{\lowercase{i}}}}\}$ is not a good description of baryonic density properties. This issue was slightly fixed by introducing another free parameter $\nu$\footnote{Note that we scales 1D baryonic density field before exponentiating in Eq. (\ref{Eq:numb_dens_baryons}) according to $\delta_{\rm B}(x,z) \rightarrow \nu \delta_{\rm B}(x,z)$, which is the approach proposed in \cite{Arya2024}.} to scale the 1D baryonic density field. However, in preliminary analysis we found that in some cases, especially when we cannot covers smaller scales ($k > 0.1$), this approach lead to the unstable results. This behaviour is demonstrated in Fig. \ref{fig:params_comp}, where we show parameter estimates obtained from the MCMC runs. These results show that the temperature at a mean density and the slope of the temperature-density relation corresponds to each other within the errors. However, in the case of $x_{\rm J}$ there is tendency to underestimate its value. 
\subsection{Evolution of the IGM Temperature}

We also compared the obtained results with the predictions of 4~UVB models from \cite{FG2020}, \cite{Haardt2012}, \cite{Puchwein2019} and the first one with rescaled {\text{H\,\textsc{\lowercase{i}}}}, {\text{He\,\textsc{\lowercase{i}}}}, {\text{He\,\textsc{\lowercase{ii}}}} photoheating rates by a factor of 0.68. As shown in \cite{Gaikwad2021}, this would results in a better match to the measured temperature evolution of the IGM at $z \geq 2$. The comparison of the IGM temperature evolution $T_{0}$ in the various models with the determined values is shown in Fig. \ref{fig:temp_evol}. For this purpose, we used our temperature evolution code {\sc{teco}} (see Appendix \ref{Appendix:TECO}).

The comparison show that our results are consistent with the ones from \cite{Boera2019} and the UVB model from \cite{FG2020}. However, the temperature at the mean density determined from the estimates \cite{Irsic2017} is higher than predicted by various model. We identify two probable reasons for such behaviour. Firstly, we used only 30 lines of sight for the MCMC analysis. The second reason, which we believe was crucial, is that the data from \cite{Boera2019} covers smaller scales. This is demonstrated in Fig.\ref{fig:comp_thermal}, where the small differences in the FPS could be found at the scales $k > 0.1$. Additionally, we do not have a way to directly compare Jeans length $x_J$ to any observations or simulations, since the lognormal uses a simplistic way to calculate baryonic density field. We found that our best-fit values of $x_J$ are of the same order as obtained in \cite{schaye_2001}, assuming Ly$\alpha$ absorbers to be in hydrostatic equilibrium at $\sim 10^4$ K. 

\begin{figure}
    \centering
    \includegraphics[width=0.75\linewidth]{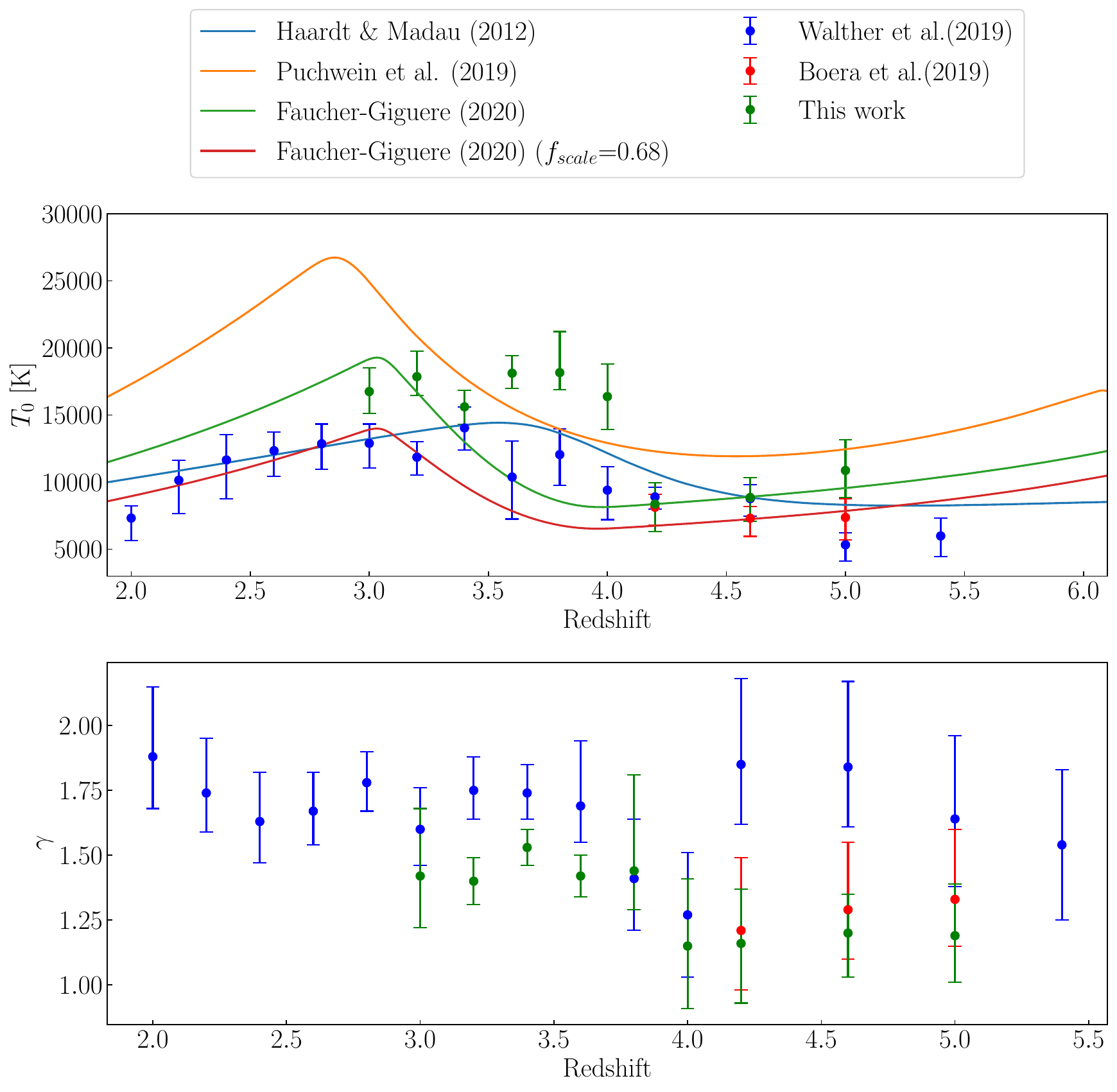}
    \caption{Comparison of the results obtained in this study with various models. We also depict values, which were determined in \cite{Boera2019} and \cite{Walther2019}.}
    \label{fig:temp_evol}
\end{figure}
\section{Conclusions}
\label{Sec:Conclusions}
In this study, we used the lognormal seminumerical simulations of~the~Lyman~$\alpha$ forest to determine the thermal parameters of the IGM. The main results could be summarized as follows:
\begin{enumerate}
    \item The lognormal simulations can effectively recover temperature at mean density, slope of the temperature-density relation and Jeans length. Thus can be used for fast exploration of parameter space of thermal parameters $T_{0}$, $\gamma$, and $x_{\rm J}$.
    \item Based on the comparison of synthetic flux power spectrum with determined one from eBOSS, it can be concluded that such an approach can be also used for the cosmological parameter inference.
    \item The synthetic FPS is consistent with the high redshift measurements of $P(k)$ from \cite{Boera2019}. However, there is the difference when comparing to the estimates from \cite{Irsic2017}, which was mainly caused by the lack of coverage on smaller scales $k > 0.1$
\end{enumerate}
These results also show that this approach is suitable for generating large number of synthetic spectra with input data and parameters. Therefore, it is  ideal for interpreting the high-quality data obtained from the QSO absorption spectra surveys. In future work, we plan to prepare analysis pipeline for measuring the FPS, which uses the lognormal seminumerical simulations.


\appendix
\section{MCMC results}
\label{Appendix:MCMC}
The parameter estimates obtained from the MCMC runs are presented in Fig. \ref{fig:MCMC}.
\begin{figure*}
    \centering
    \includegraphics[width = \textwidth]{figs/3_paramsMCMC.pdf}
    \caption{Parameter estimates obtained from the MCMC runs. Red solid line corresponds to the median value, orange and red dashed lines corresponds to the 16th and 84th percentile, respectively.}
    \label{fig:MCMC}
\end{figure*}


\section{Temperature evolution of the IGM (\sc teco)}
\label{Appendix:TECO}
\begin{figure*}
    \centering
    \includegraphics[width=\linewidth]{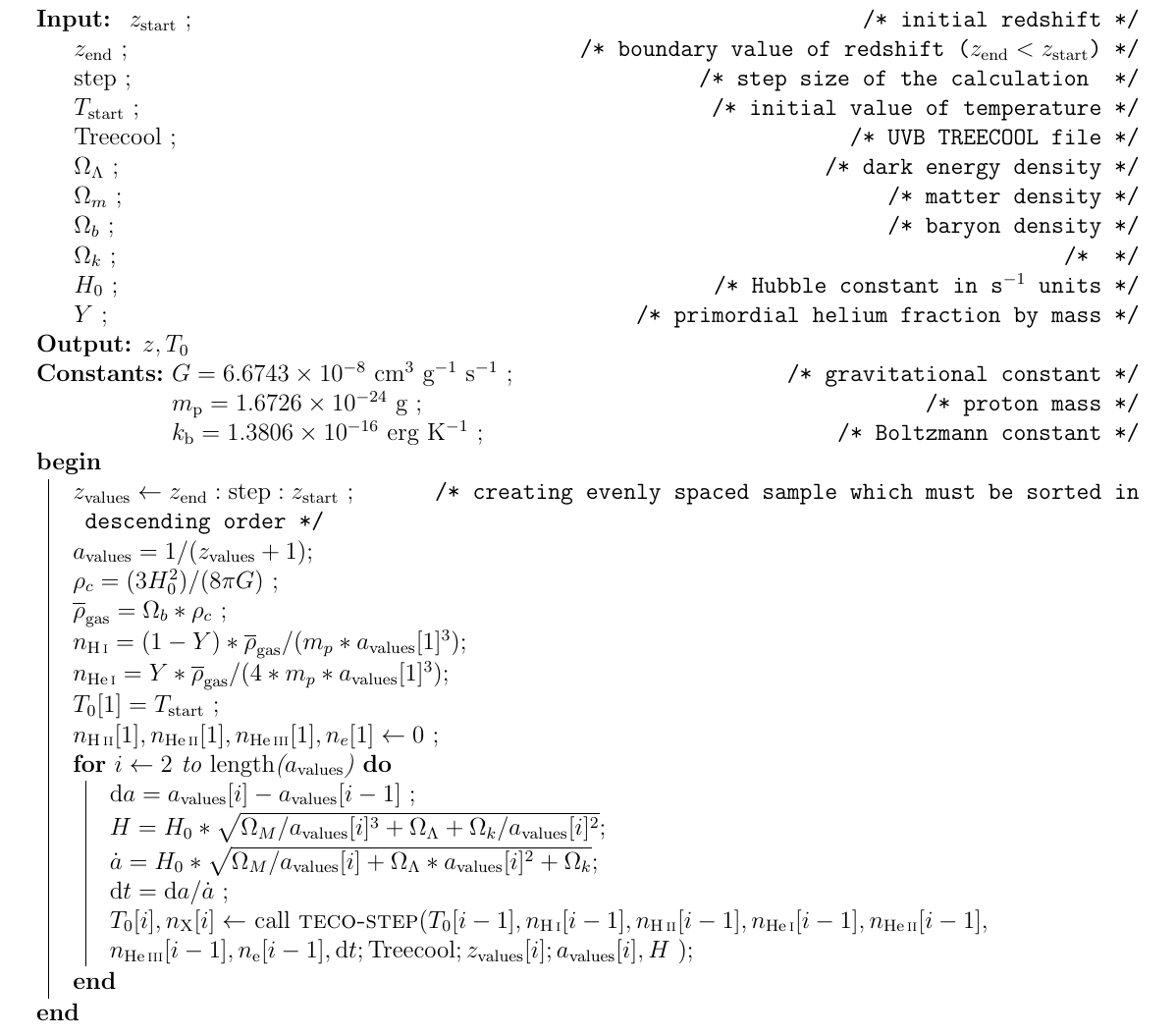}
    \caption{Algorithm of our temperature evolution code {\sc teco}.}
    \label{fig:TECO_step1}
\end{figure*}

\begin{figure*}
    \centering
    \includegraphics[width=\linewidth]{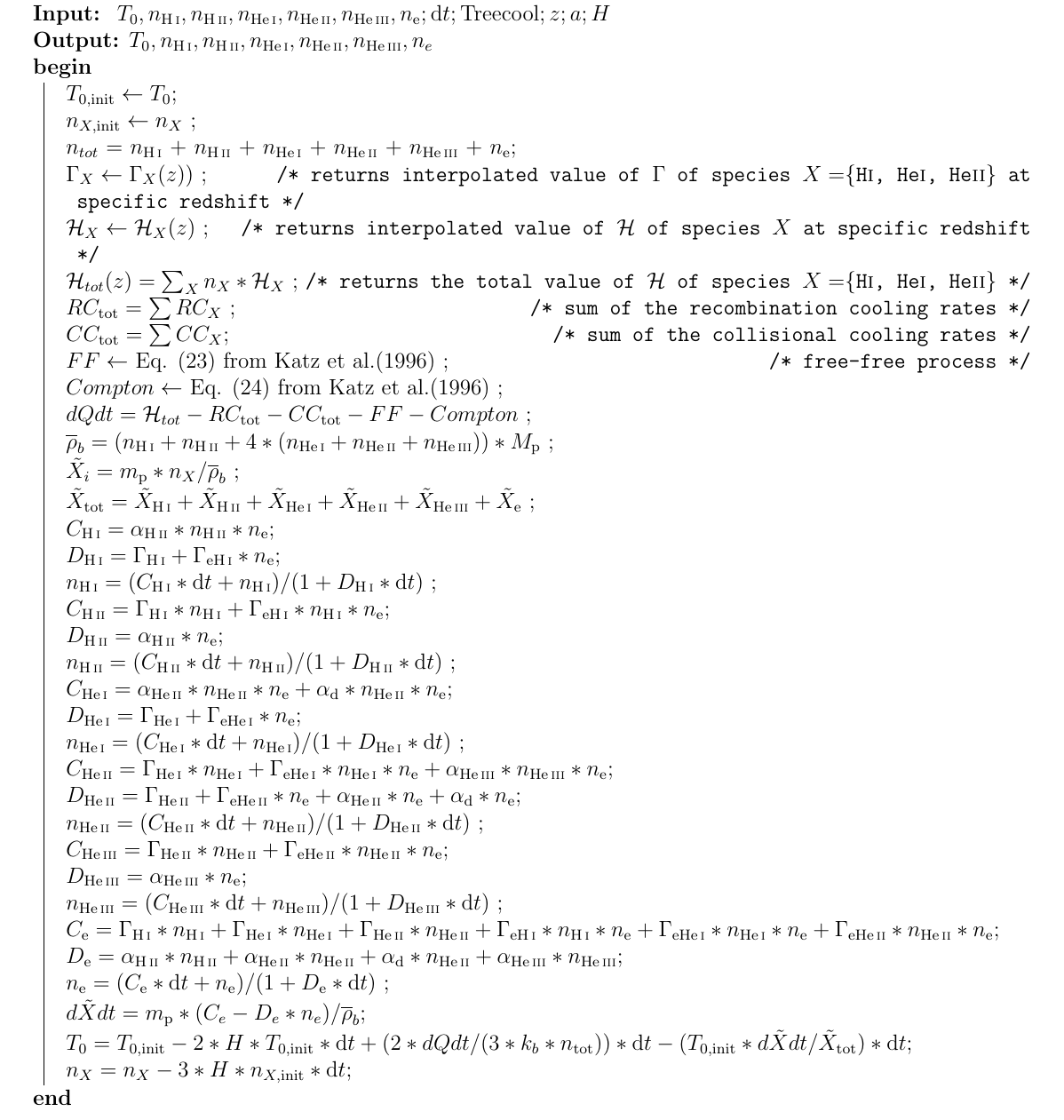}
    \caption{Algorithm of the {\sc teco-step}.}
    \label{fig:TECO_step2}
\end{figure*}

To derive the history of $T_{0}$ for various photoheating models, we solve the equation, which describe the temperature evolution of a Lagrangian fluid element at the cosmic mean density, i.e., with $\Delta = 1$ \citep{Miralda_Escude1994,HuiGnedin,McQuinn2016,Boera2019}:
\begin{equation}
    \frac{{\rm d}T}{{\rm d}t} = -2HT + \frac{2T}{3\Delta}\frac{{\rm d}\Delta}{{\rm d}t} - \frac{T}{\sum_{i}X_{i}}\frac{{\rm d}\sum_{i}X_{i}}{{\rm d}t} + \frac{2}{3k_{B}n_{\rm tot}}\frac{{\rm d}Q}{{\rm d}t},
    \label{Eq:TECO}
\end{equation}
where ${\rm d} / {\rm d}t$ is the Lagrangian derivative with respect to proper time, $H$ is the Hubble parameter and $n_{b}$ is the proper number density of all particles. The first term on the right-hand side takes into account cooling due to adiabatic expansion, while the second term gives adiabatic heating and cooling due to the structure formation. The thermal history at the mean density has been shown to depend weakly on this second term \citep{McQuinn2016}, therefore we ignored it in our calculation. The third term accounts for the change of internal energy per particle due to the change in the number of particles. The last term on the right-hand side of Eq.(\ref{Eq:TECO}) describes the effect of heating and cooling rates and can be expanded as follows:
\begin{equation}
    \frac{{\rm d}Q}{{\rm d}t} = \sum_{X} \frac{{\rm d}Q_{{\rm photo},X}}{{\rm d}t} + \frac{{\rm d}Q_{\rm Compton}}{{\rm d}t} + \sum_{i} \sum_{X} R_{i,X} n_{e} n_{X},
\end{equation}
where ${{\rm d}Q_{{\rm photo},X}}/{{\rm d}t}$ is the photoheating rate of ion $X =$ \{H{\sc i}, He{\sc i}, He{\sc ii}\}, ${{\rm d}Q_{\rm Compton}}/{{\rm d}t}$ is the Compton cooling rate, and $R_{i,X}$ is the cooling rate coefficient for the ion $X$ and cooling mechanism $i$ [see Tab.1 in \cite{Katz1996}]. Besides, in the case of  radiative cooling processes, we also include the inverse Compton cooling off the microwave background at the rate 
\begin{equation}
    \Lambda_{\rm C} = 5.41 \times 10^{-36} n_{\rm e} T (1 + z)^{4}     [\rm erg\,s^{-1}\,cm^{-3}]
\end{equation}
The number density of each species depends on the recombination and ionization rates. Therefore, we solve a kinetic network for the rate of change of species density $n_{i}$ with the general form \citep{Nagamine2018}:
\begin{equation}
    \frac{\partial n_{i}}{\partial t} = \sum_{j} \sum_{l} k_{jl} n_{j} n_{l} + \sum_{j} I_{j} n_{j},
\end{equation}
where $k_{jl}$ is the rate for reactions involving species $j$ and $l$ and $I_{j}$ is the appropriate radiative rate. This equation can be written schematically
\begin{equation}
    \frac{\partial n_{i}}{\partial t} = C_{i} (T, n_{j}) - D_{i} (T, n_{j}) n_{i},
    \label{Eq:numb_density2}
\end{equation}
where $C_{i}$ represents the total creation rate of species $i$ (given the temperature $T$ and other species densities) and $D_{i}$ is the destruction rate of the same species, which must be proportional to $n_{i}$ including both radiative and collisional processes \citep[see Tab. 2 in the study by][]{Katz1996}. Due to the reproducibility, we describe the whole algorithm of calculation in Fig. \ref{fig:TECO_step1} and \ref{fig:TECO_step2}. It is worth noting that for the calculation of the Eq. (\ref{Eq:numb_density2}) we used the backwards difference formula (BDF) due to its stability \citep{Smith2017}. Discretization of Eq.(\ref{Eq:numb_density2}) yields 
\begin{equation}
    n^{t+\Delta t} = \frac{C^{t+\Delta t} \Delta t + n^{t}}{1 + D^{t+\Delta t} \Delta t}.
    \label{BDF_step}
\end{equation}
As noted in \citep{Smith2017}, the partial forwarding updating can be done by solving the various species in a specified order and using the updated species densities in the following partial step. In this work, we used the six species model: H\,{\sc i}, H\,{\sc ii}, He\,{\sc i}, He\,{\sc ii}, He\,{\sc iii} and e$^{-}$. This order was chosen based on the study \cite{Anninos1997} and was determined through experimentations.
%

\bibliographystyle{JHEP}
\bibliography{biblio.bib}

\providecommand{\href}[2]{#2}\begingroup\raggedright\begin{thebibliography}{10}

\bibitem{viel+02}
M.~{Viel}, S.~{Matarrese}, H.J.~{Mo}, M.G.~{Haehnelt} and T.~{Theuns}, \emph{{Probing the intergalactic medium with the Ly{\ensuremath{\alpha}} forest along multiple lines of sight to distant QSOs}}, \href{https://doi.org/10.1046/j.1365-8711.2002.05060.x}{\emph{Mon. Not. R. Astron. Soc.} {\bfseries 329} (2002) 848} [\href{https://arxiv.org/abs/astro-ph/0105233}{{\ttfamily astro-ph/0105233}}].

\bibitem{Viel2002}
M.~{Viel}, S.~{Matarrese}, H.J.~{Mo}, T.~{Theuns} and M.G.~{Haehnelt}, \emph{{Modelling the IGM and the Ly {\ensuremath{\alpha}} forest at high redshift from the dark matter distribution}}, \href{https://doi.org/10.1046/j.1365-8711.2002.05803.x}{\emph{Mon. Not. R. Astron. Soc.} {\bfseries 336} (2002) 685} [\href{https://arxiv.org/abs/astro-ph/0203418}{{\ttfamily astro-ph/0203418}}].

\bibitem{demi_2011}
M.~{Demia{\'n}ski}, A.~{Doroshkevich}, S.~{Pilipenko} and S.~{Gottl{\"o}ber}, \emph{{Simulated evolution of the dark matter large-scale structure of the Universe}}, \href{https://doi.org/10.1111/j.1365-2966.2011.18265.x}{\emph{Mon. Not. R. Astron. Soc.} {\bfseries 414} (2011) 1813} [\href{https://arxiv.org/abs/1102.2951}{{\ttfamily 1102.2951}}].

\bibitem{FontRibera2012}
A.~{Font-Ribera}, P.~{McDonald} and J.~{Miralda-Escud{\'e}}, \emph{{Generating mock data sets for large-scale Lyman-{\ensuremath{\alpha}} forest correlation measurements}}, \href{https://doi.org/10.1088/1475-7516/2012/01/001}{\emph{J. Cosmol. Astropart. Phys.} {\bfseries 2012} (2012) } [\href{https://arxiv.org/abs/1108.5606}{{\ttfamily 1108.5606}}].

\bibitem{Chabanier2019}
S.~{Chabanier}, N.~{Palanque-Delabrouille}, C.~{Y{\`e}che}, J.-M.~{Le Goff}, E.~{Armengaud}, J.~{Bautista} et~al., \emph{{The one-dimensional power spectrum from the SDSS DR14 Ly{\ensuremath{\alpha}} forests}}, \href{https://doi.org/10.1088/1475-7516/2019/07/017}{\emph{J. Cosmol. Astropart. Phys.} {\bfseries 2019} (2019) 017} [\href{https://arxiv.org/abs/1812.03554}{{\ttfamily 1812.03554}}].

\bibitem{Bourboux2020}
H.~{du Mas des Bourboux}, J.~{Rich}, A.~{Font-Ribera}, V.~{de Sainte Agathe}, J.~{Farr}, T.~{Etourneau} et~al., \emph{{The Completed SDSS-IV Extended Baryon Oscillation Spectroscopic Survey: Baryon Acoustic Oscillations with Ly{\ensuremath{\alpha}} Forests}}, \href{https://doi.org/10.3847/1538-4357/abb085}{\emph{Astrophys. J.} {\bfseries 901} (2020) 153} [\href{https://arxiv.org/abs/2007.08995}{{\ttfamily 2007.08995}}].

\bibitem{Bi1992}
H.G.~{Bi}, G.~{Boerner} and Y.~{Chu}, \emph{{An alternative model for the Ly-alpha absorption forest.}}, {\emph{Astron. Astrophys.} {\bfseries 266} (1992) 1}.

\bibitem{Bi1993}
H.~{Bi}, \emph{{Lyman-Alpha Absorption Spectrum of the Primordial Intergalactic Medium}}, \href{https://doi.org/10.1086/172380}{\emph{Astrophys. J.} {\bfseries 405} (1993) 479}.

\bibitem{Bi1995}
H.~{Bi}, J.~{Ge} and L.-Z.~{Fang}, \emph{{A Simulation of LY alpha Absorption Forests in Linear Approximation of Cold Dark Matter and Cold Plus Hot Dark Matter Models}}, \href{https://doi.org/10.1086/176282}{\emph{Astrophys. J.} {\bfseries 452} (1995) 90} [\href{https://arxiv.org/abs/astro-ph/9504061}{{\ttfamily astro-ph/9504061}}].

\bibitem{Bi1997}
H.~{Bi} and A.F.~{Davidsen}, \emph{{Evolution of Structure in the Intergalactic Medium and the Nature of the Ly{\ensuremath{\alpha}} Forest}}, \href{https://doi.org/10.1086/303908}{\emph{Astrophys. J.} {\bfseries 479} (1997) 523} [\href{https://arxiv.org/abs/astro-ph/9611062}{{\ttfamily astro-ph/9611062}}].

\bibitem{petitjean_1995}
P.~{Petitjean}, J.P.~{Mueket} and R.E.~{Kates}, \emph{{The Ly{\ensuremath{\alpha}} forest at low redshift: tracing the dark matter filaments.}}, {\emph{Astron. Astrophys.} {\bfseries 295} (1995) L9} [\href{https://arxiv.org/abs/astro-ph/9502100}{{\ttfamily astro-ph/9502100}}].

\bibitem{croft2002}
R.A.C.~Croft, D.H.~Weinberg, M.~Bolte, S.~Burles, L.~Hernquist, N.~Katz et~al., \emph{Toward a precise measurement of matter clustering: Ly {\ensuremath{\alpha}} forest data at redshifts 2–4}, \href{https://doi.org/10.1086/344099}{\emph{Astrophys. J.} {\bfseries 581} (2002) 20–52}.

\bibitem{cj91}
P.~{Coles} and B.~{Jones}, \emph{{A lognormal model for the cosmological mass distribution.}}, \href{https://doi.org/10.1093/mnras/248.1.1}{\emph{Mon. Not. R. Astron. Soc.} {\bfseries 248} (1991) 1}.

\bibitem{HuiGnedin}
L.~{Hui} and N.Y.~{Gnedin}, \emph{{Equation of state of the photoionized intergalactic medium}}, \href{https://doi.org/10.1093/mnras/292.1.27}{\emph{Mon. Not. R. Astron. Soc.} {\bfseries 292} (1997) 27} [\href{https://arxiv.org/abs/astro-ph/9612232}{{\ttfamily astro-ph/9612232}}].

\bibitem{Choudhury2001}
T.R.~{Choudhury}, R.~{Srianand} and T.~{Padmanabhan}, \emph{{Semianalytic Approach to Understanding the Distribution of Neutral Hydrogen in the Universe: Comparison of Simulations with Observations}}, \href{https://doi.org/10.1086/322327}{\emph{Astrophys. J.} {\bfseries 559} (2001) 29} [\href{https://arxiv.org/abs/astro-ph/0012498}{{\ttfamily astro-ph/0012498}}].

\bibitem{Arya2023}
B.~{Arya}, T.R.~{Choudhury}, A.~{Paranjape} and P.~{Gaikwad}, \emph{{Lognormal seminumerical simulations of the Lyman {\ensuremath{\alpha}} forest: comparison with full hydrodynamic simulations}}, \href{https://doi.org/10.1093/mnras/stad386}{\emph{Mon. Not. R. Astron. Soc.} {\bfseries 520} (2023) 4023} [\href{https://arxiv.org/abs/2206.08013}{{\ttfamily 2206.08013}}].

\bibitem{Arya2024}
B.~{Arya}, T.~{Roy Choudhury}, A.~{Paranjape} and P.~{Gaikwad}, \emph{{A modified lognormal approximation of the Lyman-{\ensuremath{\alpha}} forest: comparison with full hydrodynamic simulations at 2 {\ensuremath{\leq}} z {\ensuremath{\leq}} 2.7}}, \href{https://doi.org/10.1088/1475-7516/2024/04/063}{\emph{J. Cosmol. Astropart. Phys.} {\bfseries 2024} (2024) 063} [\href{https://arxiv.org/abs/2310.12720}{{\ttfamily 2310.12720}}].

\bibitem{Boera2019}
E.~{Boera}, G.D.~{Becker}, J.S.~{Bolton} and F.~{Nasir}, \emph{{Revealing Reionization with the Thermal History of the Intergalactic Medium: New Constraints from the Ly{\ensuremath{\alpha}} Flux Power Spectrum}}, \href{https://doi.org/10.3847/1538-4357/aafee4}{\emph{Astrophys. J.} {\bfseries 872} (2019) 101} [\href{https://arxiv.org/abs/1809.06980}{{\ttfamily 1809.06980}}].

\bibitem{Walther2019}
M.~{Walther}, J.~{O{\~n}orbe}, J.F.~{Hennawi} and Z.~{Luki{\'c}}, \emph{{New Constraints on IGM Thermal Evolution from the Ly{\ensuremath{\alpha}} Forest Power Spectrum}}, \href{https://doi.org/10.3847/1538-4357/aafad1}{\emph{Astrophys. J.} {\bfseries 872} (2019) 13} [\href{https://arxiv.org/abs/1808.04367}{{\ttfamily 1808.04367}}].

\bibitem{Gaikwad2021}
P.~{Gaikwad}, R.~{Srianand}, M.G.~{Haehnelt} and T.R.~{Choudhury}, \emph{{A consistent and robust measurement of the thermal state of the IGM at 2 {\ensuremath{\leq}} z {\ensuremath{\leq}} 4 from a large sample of Ly {\ensuremath{\alpha}} forest spectra: evidence for late and rapid He II reionization}}, \href{https://doi.org/10.1093/mnras/stab2017}{\emph{Mon. Not. R. Astron. Soc.} {\bfseries 506} (2021) 4389} [\href{https://arxiv.org/abs/2009.00016}{{\ttfamily 2009.00016}}].

\bibitem{McDonald2000}
P.~{McDonald}, J.~{Miralda-Escud{\'e}}, M.~{Rauch}, W.L.W.~{Sargent}, T.A.~{Barlow}, R.~{Cen} et~al., \emph{{The Observed Probability Distribution Function, Power Spectrum, and Correlation Function of the Transmitted Flux in the Ly{\ensuremath{\alpha}} Forest}}, \href{https://doi.org/10.1086/317079}{\emph{Astrophys. J.} {\bfseries 543} (2000) 1} [\href{https://arxiv.org/abs/astro-ph/9911196}{{\ttfamily astro-ph/9911196}}].

\bibitem{Viel2004}
M.~{Viel}, J.~{Weller} and M.G.~{Haehnelt}, \emph{{Constraints on the primordial power spectrum from high-resolution Lyman {\ensuremath{\alpha}} forest spectra and WMAP}}, \href{https://doi.org/10.1111/j.1365-2966.2004.08498.x}{\emph{Mon. Not. R. Astron. Soc.} {\bfseries 355} (2004) L23} [\href{https://arxiv.org/abs/astro-ph/0407294}{{\ttfamily astro-ph/0407294}}].

\bibitem{Planck2014}
{Planck Collaboration}, P.A.R.~{Ade}, N.~{Aghanim}, C.~{Armitage-Caplan}, M.~{Arnaud}, M.~{Ashdown} et~al., \emph{{Planck 2013 results. XVI. Cosmological parameters}}, \href{https://doi.org/10.1051/0004-6361/201321591}{\emph{Astron. Astrophys.} {\bfseries 571} (2014) A16} [\href{https://arxiv.org/abs/1303.5076}{{\ttfamily 1303.5076}}].

\bibitem{Lewis2000}
A.~{Lewis}, A.~{Challinor} and A.~{Lasenby}, \emph{{Efficient Computation of Cosmic Microwave Background Anisotropies in Closed Friedmann-Robertson-Walker Models}}, \href{https://doi.org/10.1086/309179}{\emph{Astrophys. J.} {\bfseries 538} (2000) 473} [\href{https://arxiv.org/abs/astro-ph/9911177}{{\ttfamily astro-ph/9911177}}].

\bibitem{Rauch1997}
M.~{Rauch}, J.~{Miralda-Escud{\'e}}, W.L.W.~{Sargent}, T.A.~{Barlow}, D.H.~{Weinberg}, L.~{Hernquist} et~al., \emph{{The Opacity of the Ly{\ensuremath{\alpha}} Forest and Implications for {\ensuremath{\Omega}}$_{b}$ and the Ionizing Background}}, \href{https://doi.org/10.1086/304765}{\emph{Astrophys. J.} {\bfseries 489} (1997) 7} [\href{https://arxiv.org/abs/astro-ph/9612245}{{\ttfamily astro-ph/9612245}}].

\bibitem{Villasenor2021}
B.~{Villasenor}, B.~{Robertson}, P.~{Madau} and E.~{Schneider}, \emph{{Effects of Photoionization and Photoheating on Ly{\ensuremath{\alpha}} Forest Properties from Cholla Cosmological Simulations}}, \href{https://doi.org/10.3847/1538-4357/abed5a}{\emph{Astrophys. J.} {\bfseries 912} (2021) } [\href{https://arxiv.org/abs/2009.06652}{{\ttfamily 2009.06652}}].

\bibitem{Irsic2017}
V.~{Ir{\v{s}}i{\v{c}}}, M.~{Viel}, T.A.M.~{Berg}, V.~{D'Odorico}, M.G.~{Haehnelt}, S.~{Cristiani} et~al., \emph{{The Lyman {\ensuremath{\alpha}} forest power spectrum from the XQ-100 Legacy Survey}}, \href{https://doi.org/10.1093/mnras/stw3372}{\emph{Mon. Not. R. Astron. Soc.} {\bfseries 466} (2017) 4332} [\href{https://arxiv.org/abs/1702.01761}{{\ttfamily 1702.01761}}].

\bibitem{Walther2018}
M.~{Walther}, J.F.~{Hennawi}, H.~{Hiss}, J.~{O{\~n}orbe}, K.-G.~{Lee}, A.~{Rorai} et~al., \emph{{A New Precision Measurement of the Small-scale Line-of-sight Power Spectrum of the Ly{\ensuremath{\alpha}} Forest}}, \href{https://doi.org/10.3847/1538-4357/aa9c81}{\emph{Astrophys. J.} {\bfseries 852} (2018) 22} [\href{https://arxiv.org/abs/1709.07354}{{\ttfamily 1709.07354}}].

\bibitem{Villasenor2022}
B.~{Villasenor}, B.~{Robertson}, P.~{Madau} and E.~{Schneider}, \emph{{Inferring the Thermal History of the Intergalactic Medium from the Properties of the Hydrogen and Helium Ly{\ensuremath{\alpha}} Forest}}, \href{https://doi.org/10.3847/1538-4357/ac704e}{\emph{Astrophys. J.} {\bfseries 933} (2022) 59} [\href{https://arxiv.org/abs/2111.00019}{{\ttfamily 2111.00019}}].

\bibitem{Lewis2002}
A.~{Lewis} and S.~{Bridle}, \emph{{Cosmological parameters from CMB and other data: A Monte Carlo approach}}, \href{https://doi.org/10.1103/PhysRevD.66.103511}{\emph{Phys. Rev. D} {\bfseries 66} (2002) 103511} [\href{https://arxiv.org/abs/astro-ph/0205436}{{\ttfamily astro-ph/0205436}}].

\bibitem{Lewis2013}
A.~{Lewis}, \emph{{Efficient sampling of fast and slow cosmological parameters}}, \href{https://doi.org/10.1103/PhysRevD.87.103529}{\emph{Phys. Rev. D} {\bfseries 87} (2013) 103529} [\href{https://arxiv.org/abs/1304.4473}{{\ttfamily 1304.4473}}].

\bibitem{Torrado2019}
J.~{Torrado} and A.~{Lewis}, ``{Cobaya: Bayesian analysis in cosmology}.'' Astrophysics Source Code Library, record ascl:1910.019, Oct., 2019.

\bibitem{Torrado2021}
J.~{Torrado} and A.~{Lewis}, \emph{{Cobaya: code for Bayesian analysis of hierarchical physical models}}, \href{https://doi.org/10.1088/1475-7516/2021/05/057}{\emph{J. Cosmol. Astropart. Phys.} {\bfseries 2021} (2021) 057} [\href{https://arxiv.org/abs/2005.05290}{{\ttfamily 2005.05290}}].

\bibitem{Gelman1992}
A.~{Gelman} and D.B.~{Rubin}, \emph{{Inference from Iterative Simulation Using Multiple Sequences}}, \href{https://doi.org/10.1214/ss/1177011136}{\emph{Statistical Science} {\bfseries 7} (1992) 457}.

\bibitem{FG2020}
C.-A.~{Faucher-Gigu{\`e}re}, \emph{{A cosmic UV/X-ray background model update}}, \href{https://doi.org/10.1093/mnras/staa302}{\emph{Mon. Not. R. Astron. Soc.} {\bfseries 493} (2020) 1614} [\href{https://arxiv.org/abs/1903.08657}{{\ttfamily 1903.08657}}].

\bibitem{Haardt2012}
F.~{Haardt} and P.~{Madau}, \emph{{Radiative Transfer in a Clumpy Universe. IV. New Synthesis Models of the Cosmic UV/X-Ray Background}}, \href{https://doi.org/10.1088/0004-637X/746/2/125}{\emph{Astrophys. J.} {\bfseries 746} (2012) 125} [\href{https://arxiv.org/abs/2406.15237}{{\ttfamily 2406.15237}}].

\bibitem{Puchwein2019}
E.~{Puchwein}, F.~{Haardt}, M.G.~{Haehnelt} and P.~{Madau}, \emph{{Consistent modelling of the meta-galactic UV background and the thermal/ionization history of the intergalactic medium}}, \href{https://doi.org/10.1093/mnras/stz222}{\emph{Mon. Not. R. Astron. Soc.} {\bfseries 485} (2019) 47} [\href{https://arxiv.org/abs/1801.04931}{{\ttfamily 1801.04931}}].

\bibitem{schaye_2001}
J.~{Schaye}, \emph{{Explaining the Lyman-alpha forest}}, {\emph{arXiv e-prints} (2001) astro} [\href{https://arxiv.org/abs/astro-ph/0112022}{{\ttfamily astro-ph/0112022}}].

\bibitem{Miralda_Escude1994}
J.~{Miralda-Escud{\'e}} and M.J.~{Rees}, \emph{{Reionization and thermal evolution of a photoionized intergalactic medium.}}, \href{https://doi.org/10.1093/mnras/266.2.343}{\emph{Mon. Not. R. Astron. Soc.} {\bfseries 266} (1994) 343}.

\bibitem{McQuinn2016}
M.~{McQuinn} and P.R.~{Upton Sanderbeck}, \emph{{On the intergalactic temperature-density relation}}, \href{https://doi.org/10.1093/mnras/stv2675}{\emph{Mon. Not. R. Astron. Soc.} {\bfseries 456} (2016) 47} [\href{https://arxiv.org/abs/1505.07875}{{\ttfamily 1505.07875}}].

\bibitem{Katz1996}
N.~{Katz}, D.H.~{Weinberg} and L.~{Hernquist}, \emph{{Cosmological Simulations with TreeSPH}}, \href{https://doi.org/10.1086/192305}{\emph{Astrophys. J.s} {\bfseries 105} (1996) 19} [\href{https://arxiv.org/abs/astro-ph/9509107}{{\ttfamily astro-ph/9509107}}].

\bibitem{Nagamine2018}
K.~{Nagamine}, \emph{{The Encyclopedia of Cosmology. Volume 2: Numerical Simulations in Cosmology}}, World Scientific Publishing (2018), \href{https://doi.org/10.1142/9496-vol2}{10.1142/9496-vol2}.

\bibitem{Smith2017}
B.D.~{Smith}, G.L.~{Bryan}, S.C.O.~{Glover}, N.J.~{Goldbaum}, M.J.~{Turk}, J.~{Regan} et~al., \emph{{GRACKLE: a chemistry and cooling library for astrophysics}}, \href{https://doi.org/10.1093/mnras/stw3291}{\emph{Mon. Not. R. Astron. Soc.} {\bfseries 466} (2017) 2217} [\href{https://arxiv.org/abs/1610.09591}{{\ttfamily 1610.09591}}].

\bibitem{Anninos1997}
P.~Anninos, Y.~Zhang, T.~Abel and M.L.~Norman, \emph{Cosmological hydrodynamics with multi-species chemistry and nonequilibrium ionization and cooling}, \href{https://doi.org/https://doi.org/10.1016/S1384-1076(97)00009-2}{\emph{New Astronomy} {\bfseries 2} (1997) 209} [\href{https://arxiv.org/abs/astro-ph/9608041}{{\ttfamily astro-ph/9608041}}].

\end{thebibliography}\endgroup






\end{document}